\documentstyle[aps,epsf,twocolumn]{revtex}

\def\myfigure#1#2{{\leftskip=0.000753\textwidth \rightskip\leftskip\small
\begin{figure}\baselineskip=14pt plus 2pt minus 1pt
\centerline{#1}\nobreak\smallskip\nobreak #2\end{figure}}}
\global\firstfigfalse

\begin{document}
\twocolumn[
\hsize\textwidth\columnwidth\hsize\csname @twocolumnfalse\endcsname

\title{Low temperature hcp to monoclinic structural transition in solid
C$_{70}$ : Ephemeral nature of the intermediate phase}

\author{G. Ghosh, V. S. Sastry, C. S. Sundar, Surajit Sengupta and T. S.
Radhakrishnan}

\address{Materials Science Division, Indira Gandhi Centre for Atomic
Research, Kalpakkam 603 102, India.}

\maketitle
\begin{abstract}
We follow the structural transformation in solid C$_{70}$ from
the high temperature hcp to a low temperature monoclinic phase
using detailed x-ray diffraction studies at controlled
cooling-rates from 0.0033 to 0.42 K/min.  Rapid cooling of the
sample gives the signature of a two-step transformation which
disappears when the system is transformed quasi-statically.
These experimental results can be rationalized using a simple
mean field, Langevin dynamical theory using a free energy
functional with minima corresponding to the parent and two
competing product phases such that one of these product phases
remains metastable throughout. The implication of our results on
the existence of the intermediate phase in the sequence,
hcp-dhcp- monoclinic, of structural transitions in solid
C$_{70}$ with the lowering of temperature is discussed.
\end{abstract}
\narrowtext]
	Following the pioneering work of Heiney {\em et
al.},\cite{HEI} on relating the structural transition at low
temperature (260 K) in solid C$_{60}$ to the development of
orientational correlations between rotating C$_{60}$ molecules,
there have been extensive studies on fullerene molecular
crystals, both solid C$_{60}$ and C$_{70}$. However, unlike in
solid C$_{60}$, the low temperature structural transitions in
the next higher fullerene C$_{70}$ are not unambiguously
established. The situation is exacerbated by the complicated
orientational dynamics of the lower symmetry (D$_{5h}$)
molecule, associated sluggishness of the transition and
complications due to intercalated solvent effects etc.  Solid
C$_{70}$ is observed to crystallize in both fcc and ideal hcp
structures, corresponding to close packing of rapidly rotating
spherically averaged molecules. The fcc solid
C$_{70}$ is first seen to
transform at $\approx$280 K to a rhombohedral structure and
subsequently to a monoclinic structure at 200 K\cite{CRIST} with the 
lowering of temperature;
both these transitions are associated with considerable
hysteresis of $\approx$ 50 K. A widely accepted viewpoint on the nature of
structural transitions in hcp solid C$_{70}$ involves two
transitions : One from the ideal hcp ({\it c}/{\it a} $\approx$
1.63) to the deformed hcp ({\it c}/{\it a} $\approx$ 1.82) at
337 K (dhcp) and a second transition to the monoclinic structure
at 276 K.\cite{GBMV,MAV,TEND} Data from differential thermal
analysis (DTA)\cite{TANA}, differential scanning calorimetry
(DSC)\cite{SWOR} and dilatometry measurement\cite{MEIN}
also lend support to this scenario. The occurence of a two-step
transition is rationalized in terms of successive freezing of
molecular orientations, first along the short C$_2$ axis and
then along the long C$_5$ axis; the structural transitions being
produced by the coupling  of these orientational degrees of
freedom to the lattice strain.

Several experiments, however, see a single-step transition from hcp to
monoclinic phase, and there are others \cite{RAMASESHA} that have reported 
three structural
transitions with the variation of temperature. For example,
Dennis {\em et al.}\cite{DEN} observed a single-step
orientational ordering of C$_{70}$ molecules, over the
temperature range of 270 to 160 K in their muon spin resonance
($\mu$SR) experiments. In our  earlier work\cite{GG1},
using high quality samples prepared by the solution route we
observed (1) the equilibrium room temperature structure of solid
C$_{70}$ to be ideal hcp and (2) a broad (270 to 170 K)
single-step transition from the ideal hcp to a monoclinic
structure implying a simultaneous freezing of all the
orientational degrees of freedom.  It is of interest to know if
the differences in the structural sequence are related to the
different (often uncontrolled) cooling-rates employed in the
earlier experiments.  If an {\em
equilibrium} intermediate phase actually exists then it would be
seen during controlled slow cooling of the system from the ideal
hcp structure.

	With a view to investigate this, x-ray diffraction (XRD)
experiments at various controlled cooling-rates (varying over
two orders of magnitude) have been carried out, and the
evolution of the monoclinic phase at low temperature followed.
Contrary to our initial expectations, though the transformation
proceeds through a single-step (starting at $\approx$270 K) when
a slow cooling-rate (0.0033 K/min.) is used,it has a two-step
sequence on cooling rapidly (0.42 K/min.). Our experimental
results (Fig. 2.) strongly suggests that the intermediate phase
reported earlier\cite{GBMV,MAV,TEND} is likely to be an artifact
of the kinetics of the transformation process and {\em not} a
true equilibrium phase.  We show that qualitative aspects of our
data for the temparature evolution of the fraction of the retained hcp
phase during the hcp to monoclinic transition can be understood
within a simple phenomenological kinetic theory employing a
general free energy functional with minima corresponding
to the parent and two product phases such that one of the
product phases is always metastable.

	Chromatographically separated C$_{70}$ powder was heated
in vacuum at 250$^o$ C for 24 hours to get rid of trapped
solvents. The x-ray diffraction (XRD) study at low temperatures
was carried out using a Siemens D-500 powder diffractometer with
a continuous flow cryostat. The sample holder was a (911) cut
silicon wafer,\cite{GVNR1} attached to the cold tip of the
cryostat using a thin layer of GE vernish. The sample
displacement errors due to thermal contraction were corrected by
simultaneously using a small quantity of NIST 640a silicon
powder as internal d-spacing standard.  The XRD patterns were
recorded at 10 K steps over the temperature range of 300 to 77
K. The sample was cooled at controlled rates. Experiments were
carried out at four different cooling-rates varying over two
orders of magnitude, {\it viz.} 0.0033 K/min., 0.22 K/min., 0.31
K/min., and 0.42 K/min. Before commencing a fresh cooling cycle
we carried out a detailed XRD run at room temperature to ensure
that the sample had regained its original hcp structure.
\myfigure{\epsfysize3.0in\epsfbox{fig1.eps}}
{\vskip0.0in Fig. \,1~~
XRD patterns showing the temperature evolution of the monoclinic
phase from the parent hcp at different cooling rates.  (a)
0.0033 K/min.;  the lowest cooling-rate employed in this
experiment (b) 0.22 K/min. (c) 0.31 K/min. and (d) 0.42 K/min.}

	As shown in Fig.1 (a), the room temperature XRD pattern
can be indexed to the ideal hcp with lattice parameters {\it a}
= 10.593 $\AA~$, {\it c} = 17.262 $\AA~$, and space group {\it
P6}$_3$/{\it mmc}\cite{GVNR2}. The low temperature pattern can
be indexed to a monoclinic structure, space group {\it P2}$_1$/{\it m},
with lattice parameters {\it a}= 10.99 $\AA~$, {\it b}= 16.16 $\AA~$,
{\it c}=9.85 $\AA~$ , $\beta $ = 107.75$^o$. At a slow cooling rate
of 0.0033 K/min, it is seen that the hcp phase transforms to the
monoclinic structure without the intervening dhcp phase,
consistent with the results reported earlier\cite{DEN}. At higher
cooling rates of 0.22, 0.31 and 0.42 K/min, XRD patterns in the
2$\theta$ range of 16$^o$ and 18$^o$, corresponding to the most
intense peaks of the hcp (110) and the monoclinic ($\bar2$11)
phases\cite{GG1}, have been measured, and these results are
shown in Figs.1 (b)-(d).

The change in the fraction of the hcp phase with temperature
has been calculated from these patterns as the area ratio of
the 100$\%$ hcp (110) reflection to the sum of the areas of the
(110) reflection of the hcp and ($\bar2$11) reflection of the
monoclinic phase, and these results are plotted in Fig.2. It is
seen that at a cooling-rate 0.0033 K/min. the hcp fraction
decreases in a single-step, within a temperature range  of about
100 K. The structural change commences at $\approx$270 K, and by
$\approx$170 K the transformation to the monoclinic phase is
complete. While apriori, such a broad transition could be attributed to
impurity effects (presence of intercalants), the observation of a
$\approx$80 K co-existence region even in the sublimed sample \cite{CRIST}
suggests that this feature is intrinsic. It is seen from Fig. 2 that the
width of the transition increases with the increase in cooling rate (e.g.,
0.22 K/min.) implying that the width is related to the sluggishness of
transition in solid C$_{70}$. The kinetic model proposed in this paper
(see below) accounts for the broad hcp to monoclinic transition solid
C$_{70}$ ({\em cf.} Fig. 2 and Fig. 3). 

The interesting aspect of Fig. 2 is that with increasing the
cooling-rate further,
one obtains a step in the transformation region, which can be seen
as a distinct plateau for the cooling-rate of 0.42 K/min, the
fastest controlled cooling rate possible with our present
experimental setup.
\myfigure{\epsfysize2.5in\epsfbox{fig2.eps}}
{\vskip0.0in Fig. \,2~~
Change in the fraction of the retained parent  hcp phase with
temperature at different cooling-rates as indicated.}

We argue below that this plateau is a
signature of the formation of an intermediate dhcp phase during  the
hcp to monoclinic transition. The dhcp phase has been reported
\cite{MEIN} to have lattice parameters of {\it a} = 10.11
$\AA~$, {\it c} = 18.50 $\AA~$. The (103) reflection of this
structure occurs at 17.59, very close to the ($\bar2$11) line of the
monoclinc phase occuring at 17.61, and has an intensity that is
only $\approx$ 30 $\%$ intensity of the strongest (110)
reflection \cite{MEIN}. Thus, the formation of the dhcp phase
will result in a decrease (increase) in the estimated monoclinic
(hcp) fraction, leading to the plateau. 
The interesting aspect of Fig.2.  is that this plateau,
corresponding to the dhcp phase, is seen during fast
cool. If the intermediate dhcp phase were an equilibrium phase,
we should have observed it in the slow cool experiments. This
suggests that intermediate phase is metastable, and
the feature seen in Fig.2 is a reflection of kinetic effects - a
model for which is developed as follows.

Being a large and somewhat asymetric molecule C$_{70}$,
possesses a rich energy landscape with many closely competing
local minima with the possiblity of the system getting trapped in
metastable phases influencing the equilibrium phase transitions
in this system. A Landau theory for the hcp - monoclinic
transition in C$_{70}$ can, in principle, be constructed along
the lines worked out in Refs\cite{HARI} for the fcc -
rhombohedral transition. In this case, the free energy
would have three nonequivalent minima corresponding to the three
(hcp, dhcp and monoclinic) phases described above and is
expected to have either one (hcp-monoclinic)  or a sequence of
two (hcp-dhcp-monoclinic) first order transitions depending on
the relative stability of the dhcp and the monoclinic phases.
The question of relative stability is, unfortunately, tricky
since it depends on the actual magnitudes of the various
coefficients in the Landau expansion. The uncertainty in the
values of these coefficients, which have to be obtained
experimentally or from accurate microscopic calculations, makes
it difficult in practice to settle this question convincingly.
\myfigure{\epsfysize2.5in\epsfbox{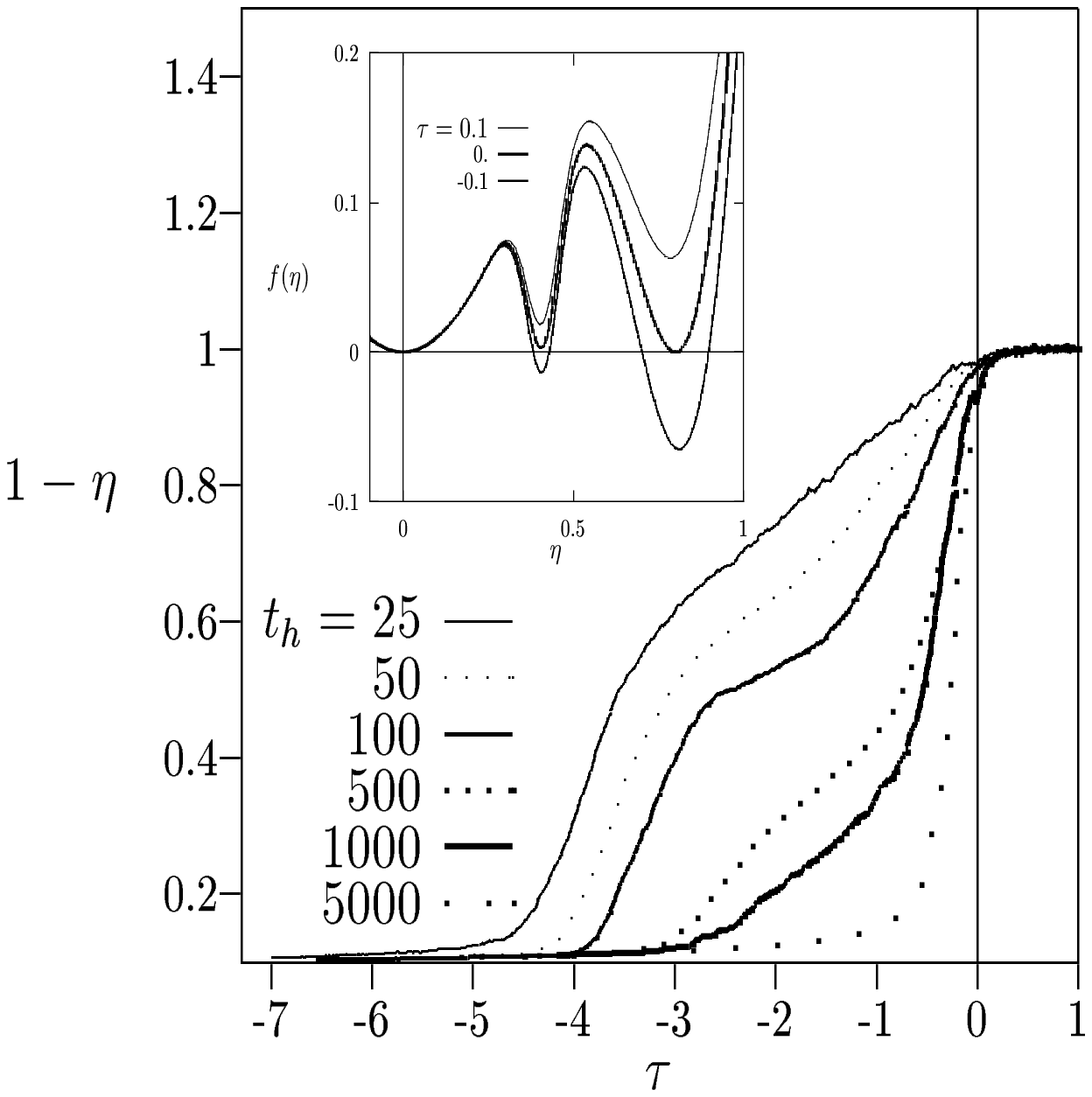}}
{\vskip0.0in Fig. \,3~~
The time and noise averaged order parameter 1-$\eta$ plotted as
a function of the dimensionless temperature $\tau$ for various
cooling rates obtained by holding the system for $t_h$ steps as
$\tau$ is reduced from $\tau=1$ in steps of $.01$. The order
parameter $\eta$ is obtained by numerically solving the
stochastic differential differential equation Eq.3. (inset) Plot
of the free energy $f(\eta)$ as a function of the order
parameter $\eta$ for $\tau = .1,0$ and $-.1$. Note that the
intermediate minimum at $\eta=0.4$ is always metastable.}

We therefore choose a simple, phenomenological approach to
understand the {\em qualitative} features of the kinetics of the
hcp-monoclinic transition as depicted in Fig. 2. without trying
to derive them with any amount of numerical accuracy. Consider a
general free energy $f(\eta)$ with three minima obtained by adding
inverted Gaussians at $\eta$=$\eta_1$ and $\eta_2$ with widths $\alpha1$
and $\alpha2$ representing the product phases, to a parabolic potential
centred at $\eta = 0$, representing the parent phase.

\begin{equation}
f = \eta^2\lbrace1-(1-\tau)(\delta e^{-\alpha_1(\eta-\eta_1)^2}+e^{-\alpha_2(\eta-\eta_2)^2})\rbrace
\label{FREE}
\end{equation}
The depth of these minima relative to the one at
$\eta=0$ is determined by the dimensionless ``relative
temperature'' $\tau \propto (T-T_c)/T_c$ (defined for $\tau < 1
$ with T$_c$ the equilibrium transition temperature)  and
$\delta$ controls the relative stability of the products for
$\eta_1$ and $\eta_2$; we have chosen $\delta$ to be such that
the minimum at $\eta_1$ is always metastable throughout the
range of $\tau$ considered. At $\tau=0$  we have a single first
order transition from the parent to the product corresponding to
$\eta=\eta_2$.  Using the (arbitrary) set of parameters
$\eta_1=0.4, \alpha_1 = 200, \eta_2=0.8, \alpha_2 = 10$ and
$\delta = 0.78$ we have plotted the free energy $f(\eta)$ for
$\tau = .1, 0$ and $-.1$ in Fig. 3 (inset).  Note that the
parameter $\delta$ has been so chosen that the intermediate
minimum is always slightly metastable compared to the other two
minima.  While our order parameter $\eta$ has no direct
connection to actual order parameters for orientational
transitions of C$_{70}$ it may be thought of as a ``distance''
along a ``minimum free energy'' path connecting the parent
minima to the two nonequivalent product minima in the true high
dimensional order parameter space.

We now investigate the dynamics of this transition and
especially the effect of the metastable minimum at
$\eta=\eta_1$.  The dynamics is governed by a Langevin
equation\cite{CHAI} for the order parameter $\eta$ (we neglect
all spatial fluctuations) evolving in the free energy landscape
$f(\eta)$ and is given by,

\begin{equation}
d\eta / dt = -\Gamma d f(\eta) /d \eta + \xi(t).
\label{TDGL}
\end{equation}
The friction coefficient $\Gamma$ is related to the variance of
the Gaussian white noise $\gamma$ by the usual
fluctuation-dissipation relation $\gamma = <\xi(t)\xi(t^\prime)>
= 2 k_B T \Gamma \delta (t-t^\prime)$, where $k_B$ is the
Boltzmann constant and $T$ the actual temperature. The value of
$\Gamma$ can, in principle, be obtained from the libron
lineshape obtained from Raman \cite{LOOS} and neutron \cite{KOSMAS}scattering
experiments.  The angular brackets signifies an average over
various realizations of the random noise.
\myfigure{\epsfysize2.5in\epsfbox{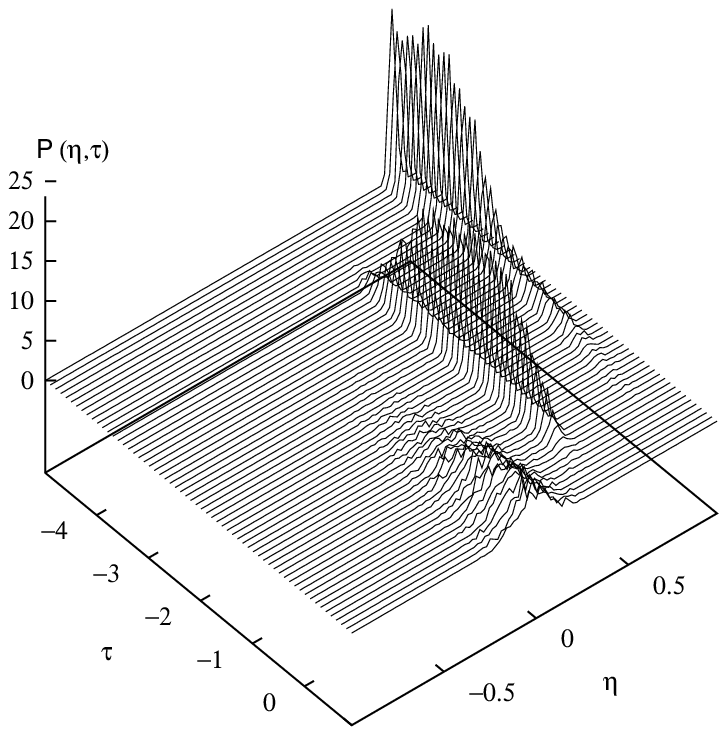}}
{\vskip0.0in Fig. \,4~~
The normalized probability distribution $P(\eta,\tau(t))$ of the
order parameter $\eta$ as a function of $\eta$ as $\tau$ is
reduced from $\tau = 1$ with a holding time of $t_h = 100$ steps
(curve number 3 in Fig. 3)} 

We simulate Eq.\ref{TDGL} using a straightforward Euler
discretization scheme with a timestep of $\Delta t= .01$. The
dimensionless temperature $\tau$ is reduced from $1$ in steps of
$.01$.  At each temperature step the system was held for $t_h$
steps (a larger value of $t_h$ implies slower cooling). The
order parameter $\eta$  averaged over $t_h$ steps and over
various realizations of the random noise (typically $300-700$)
is plotted as a function of $\tau$ in Fig. 3 (for the above set
of parameters for $f(\eta)$ and $\gamma = 1.8$). The resemblance
with the cooling curves of Fig. 2 is remarkable. Both extremely
fast and slow cooling obtains a single step transformation ---
the transformation becoming sharper as the cooling rate is
reduced.  For intermediate values of the cooling rate a plateau
develops.  The plateau results when the system is trapped,
temporarily, in the intermediate phase as it evolves from the
parent to the product structure.  This can be illustrated as
follows.

In Fig.4 we have plotted the
probability distributions $P(\eta,\tau(t))$ as a function of
$\tau(t)$ for a collection of $500$ independent systems cooled
from $\tau = 1$ to $\tau = -4.0$ with $t_h = 100 \Delta t$.
Starting from a delta function at the origin at $t=0$ the
$P(\eta,t)$ spreads out with time and eventually evolves into a
strong peak centered at the equilibrium value of $\eta$.
However, we observe that the system spends a considerable amount
of time in the secondary metastable minimum near $\eta_1$ which
roughly corresponds to the plateau observed in Fig.3 (curve 3).
In other words, at slow cooling rates 
the system gets enough time to relax to the
global minimum. For very fast quenches, on the other hand, by the 
time the system relaxes from the parent to the
metastable intermediate phase, $\tau$ decreases to the extent
that the barriers are no longer sufficient to trap it there.
Similar qualitative behaviour is seen over a broad range of
values for our parameters as long as the intermediate minimum
lies sufficiently close to the true global minimum and it is
sharp enough to trap the system effectively. This behaviour is
quite general and is analogous to that in
ferroelectrics where a similar kinetic stabilization of the
metastable paraelectric state produces double hysteresis
loops\cite{HYS}.

To paraphrase, the low temperature structural transformation of
solid C$_{70}$ exhibits interesting kinetic effects, as
evidenced by XRD experiments carried out at various cooling
rates. At extremely slow cooling rates (0.0033K/min), rarely
employed in the studies so far, the transformation to the
monoclinic phase proceeds through a single step. This implies
that the equilibrium transformation corresponds to the
simultaneous freezing of all orienational degrees of freedom,
and recently \cite{GG2} a structural model for the low
temperature monoclinic structure has been proposed. At high
cooling rate the transformation from ideal hcp to monoclinic
phase proceeds through an intermediate phase. With support from
a theoretical model, we show that such a situation
can arise if the intermediate phase is metastable.  In the light
of present investigations theoretical calculations on the
stabilty of dhcp phase need to be re-examined.

\acknowledgments
	We gratefully acknowledge Y. Hariharan and  A. Bharathi
for providing good quality C$_{70}$ powder. We thank K.P.N.
Murthy and Madan Rao for useful discussions.

\end{document}